# On solution of Klein-Gordon equation in scalar and vector potentials

Liu Changshi

Physics division, Department of mechanical and electrical engineering, Jiaxing College, Zhejiang, 314001, P. R. China

Liucs4976@sohu.com

Abstract

Based on the Coulomb gauge, the accurate Klein-Gordon equation in static scalar and vector potentials was derived from Klein-Gordon equation in electromagnetic environment. The correct equation developed in this comment demonstrates that so-called the Klein-Gordon equation with scalar and vector potentials is incorrect; therefore, some papers published to solve Klein-Gordon equation with equal scalar and vector potential are also wrong.

Keywords: Accurate Klein-Gordon equation; Scalar potential; Vector potential; Derive; Solution.

PACS：03.00.00， 03.65.-w

1. Introduction

The description of phenomena at higher energy requires the investigation of relativistic wave equation. This means equation which is invariant under Lorentz transformation. The presence of strong fields introduces quantum phenomena, such as super-criticality and spontaneous pair production that can not be described using existing techniques. Perhaps because of the difficulties on mathematics, in relativistic quantum mechanics only a few bound state solutions of the Klein-Gordon equation with scalar and vector potentials can be solved [1]. In the last twenty years, a lot of attentions have been paid to solve these relativistic energy levels and wave function for various quantum systems [2-27]. However, all of these works in this area are based on equation which is expressed as [28]

($\hbar = c = 1$, Natural unit)

$$\left\{ \frac{d^2}{dr^2} - \frac{(l+1)l}{r^2} + \left[E - V(r)\right]^2 - \left[m_0 + S(r)\right]^2 \right\} u(r) = 0 \quad (1)$$

Where vector potential is $V(r)$ and $S(r)$ is scalar potential. However, up to now, there has hardly been one who checked whether equation (1) is absolute correct or not. Unfortunately, it will be shown in this paper that nobody can derive equation (1) from Klein-Gordon equation with an electromagnetic field. Therefore, it is interesting to know why there have been so many researches that used equation (1) to detect Klein-Gordon equation; the reason for this question is that they have misunderstood one passage in one textbook [29],

$$[c\hat{\vec{\alpha}} \cdot \hat{\vec{p}} + \hat{\beta}(m_0 c^2 + V_2) - (E - V_1)]\Psi = 0 \quad (2)$$

Obviously, equation (2) is relative invariant, but equation (1) is not relative invariant.

2. Basic equation

With the minimal coupling, the free Klein-Gordon equation was transformed into the free Klein-Gordon equation with electromagnetic field [1, 30],

$$\frac{1}{c^2}\left(i\hbar\frac{\partial}{\partial t} - eA_0\right)^2 \Psi(\vec{r},t) = \left[\left(i\hbar\nabla + \frac{e}{c}\vec{A}\right)^2 + m_0^2 c^2\right]\Psi \quad (3)$$

Where $m_0$ is rest mass, equation (3) is invariable under Lorentz transformation. When both electronic field and magnetic field is static, a stationary state of the Klein-Gordon equation has the form

$$\Psi(\vec{r},t) = \Psi(\vec{r})\exp\left(-i\frac{E}{\hbar}t\right) \quad (4)$$

Where $|E|$ is the energy of the system.

$$\frac{1}{c^2}(E - eA_0)^2 \Psi(\vec{r}) = \left[\left(i\hbar\nabla + \frac{e}{c}\vec{A}\right)^2 + m_0^2 c^2\right]\Psi \quad (5)$$

It is not difficultly to develop equation (3) into [31]

$$\frac{1}{c^2}(E - eA_0)^2 \Psi(\vec{r}) = \left[-\hbar^2\Delta + i\hbar\frac{e}{c}(\nabla \cdot \vec{A}) + i\hbar\frac{e}{c}\vec{A}\cdot\nabla + \left(\frac{e}{c}\vec{A}\right)^2 + m_0^2 c^2\right]\Psi$$

(6)

$$[\left(\frac{E-eA_0}{\hbar c}\right)^2 + \Delta - \left(\frac{e}{\hbar c}\vec{A}\right)^2 - \frac{m_0^2 c^4}{\hbar^2 c^2}]\Psi(\vec{r})$$
$$+i\hbar\frac{e}{c}\left(\nabla\cdot\vec{A}+\vec{A}\cdot\nabla\right)\Psi(\vec{r})=0 \tag{7}$$

Static magnetic field always preserves the Coulomb gauge and this meaning can be expressed by

$$\nabla\cdot\vec{A}=0 \tag{8}$$

Introducing this expression into equation (7) yields

$$[(\frac{E-eA_0}{\hbar c})^2 + \Delta - (\frac{e}{\hbar c}\vec{A})^2 - \frac{m_0^2 c^4}{\hbar^2 c^2} + i\hbar\frac{e}{c}\vec{A}\cdot\nabla]\Psi(\vec{r})=0 \tag{9}$$

Equation (9) is customarily called the general form of Klein-Gordon equation with static electromagnetic field.

The author of ref. [28] neither told reader who had calculated equation (1) nor derived this equation by himself, even if there are not any approximations in Ref. [28] to obtain Eq. (1); therefore, reader can not know how to obtain equation (1) in ref. [28]. It is easy for anyone who master quantum mechanics to recognize that the finally result of equation (1) is

$$\left\{\frac{d^2}{dr^2} - \frac{l(l+1)}{r^2} + E^2 - m_0^2 + V^2 - S^2 - 2(EV + m_0 S)\right\}u(r) = 0 \tag{10}$$

Because equation (9) is one partial differential equation in form of complex, there is no reason to say that the real part of equation (9) is zero, meanwhile, the imaginary part of equation (9) can not equal to zero, and therefore, equation (10) can not be derived from equation (9). Because another form of equation (1) expressed by equation (10) can not be derived from equation (9) which is Klein-Gordon equation in static scalar and vector potential, equation (1) is incorrect. Moreover, some papers [2-27] which are so-called solution of Klein-Gordon equation in equal scalar and vector potentials are wrong too.